\documentstyle[preprint,aps,12pt]{revtex}
\tightenlines
\begin{document}
\draft
%\hfill\vbox{\baselineskip14pt
%            \hbox{\bf KEK-TH-545}
%            \hbox{\bf KEK Preprint 97-159}
%            \hbox{September 1997}}
%            \hbox{\today}}
\baselineskip20pt
\vskip 0.2cm 
\begin{center}
{\Large\bf Proposal of a Topological M(atrix) Theory}
\end{center} 
\vskip 0.2cm 
\begin{center}
\large S.~Alam 
\end{center}
\begin{center}
%{\it Theory Group, KEK, Tsukuba, Ibaraki 305, Japan }
{\it Physical Science Division, ETL, Tsukuba, Ibaraki 305, Japan }
\end{center}
%%%%%%%%%%%%%%%%%%%%%%%%%%%%%%%%%%%%%%
\vskip 0.2cm 
\begin{center} 
\large Abstract
\end{center}
\begin{center}
\begin{minipage}{14cm}
\baselineskip=18pt
\noindent
%%%%%%%%%%%%%%%%%%%%%%%%%%%%Abstract%%%%%%%%%%%%%%%%%%%%
%\input IAabs.tex

Keeping in mind the several models of M(atrix)
theory we attempt to understand the possible structure
of the topological M(atrix) theory ``underlying'' these
approaches. In particular we are motivated by the issue
about the nature of the structure of the vacuum of the 
topological M(atrix) theory and how this could be related 
to the vacuum of the electroweak theory. In doing so
we are led to a simple topological matrix model. 
Moreover it is intuitively expected from the current 
understanding that the
noncommutative nature of ``spacetime'' and background
independence should lead to a topological Model. 
%%%%%%%%%%%%%%%%%%%%%%%%%%%%%%%%%%%%%%%%%%%%%%%%%%%%
%Moreover it is expected from the current understanding that the
%noncommutative nature of ``spacetime'' and background
%independence should lead to a Topological Model. 
%%%%%%%%%%%%%%%%%%%%%%%%%%%%%%
The main purpose of this note is to propose a simple 
topological matrix model which bears relation
to F and M theories. Suggestions on the origin of
the chemical potential term appearing in the matrix
models are given. 

%In an attempt to understand the possible structure
%of the topological M(atrix) theory ``underlying'' the
%several approaches[models] to[of] M(atrix)
%theory.

%\input IAFIG1.tex
\end{minipage}
\end{center}
\vfill

\baselineskip=20pt
\normalsize

\newpage
\setcounter{page}{2}
%\input IAsec1.tex
%%%%%%%%%%%%%%%%%%%%%%%%%body-of-paper%%%%%%%%%%%%%%%%%%%%%%%%%
%Ins
%last modified 20/12/97
%IAsec1.tex
\section{Introduction}
By starting with the Green-Schwarz action for type
IIB superstring and considering its path integral
in the Schild gauge Ishibashi et al.~\cite{Kaw97} 
proposed a matrix model action, 
\begin{equation}
S=-\alpha\left(\frac{1}{4}
Tr([A_{\mu},A_{\nu}]^{2})
+\frac{1}{2}Tr(\overline{\psi}\Gamma^{\mu}
[A_{\mu},\psi])\right)+\beta N.
%Tr([A_{\mu},A_{\nu}][A^{\mu},A^{\nu}])
%\alpha=\frac{1}{g^2}
\label{F1}
\end{equation}
$\psi$ is a ten-dimensional Majorana-Weyl spinor
field, and $A_{\mu}$ and $\psi$ are $N \times N$
Hermitian matrices. The action in Eq.~\ref{F1} after
dropping the term proportional to $N$ [chemical
potential term] constitutes a large-N reduced model 
of the ten-dimensional super Yang-Mills theory. 
By noting that $N$ in Eq.~\ref{F1} is a dynamical variable
Fayyazuddin et al.~\cite{Fay97} proposed a slightly
general form of Eq.~\ref{F1}, viz
\begin{equation}
S=-\alpha\left(\frac{1}{4}
Tr(Y^{-1}[A_{\mu},A_{\nu}]^{2})
+\frac{1}{2}Tr(\overline{\psi}\Gamma^{\mu}
[A_{\mu},\psi])\right)+\beta Tr Y,
%Tr([A_{\mu},A_{\nu}][A^{\mu},A^{\nu}])
%\alpha=\frac{1}{g^2}
\label{F2}
\end{equation}
We note that $N$ in the last term in Eq.~\ref{F1} is 
the $N \times N$ identity matrix, i.e. $Tr I= N $.
The positive definite Hermitian matrix $Y^{ij}$
in Eq.~\ref{F2} is a dynamical variable with its origin
in the $\sqrt{g}$ appearing in the Schild action 
\cite{Kaw97} 
\begin{equation}
S_{_{\rm Schild}}=\int d^2\sigma \left(\alpha\sqrt{g}
(\frac{1}{4}\{X_{\mu},X_{\nu}\}^{2}_{_{PB}}
-\frac{i}{2}\overline{\psi}\Gamma^{\mu}
\{X_{\mu},\psi\}_{_{PB}})+\beta \sqrt{g}\right),
\label{F3}
\end{equation}
in the notation of \cite{Kaw97}.

The modification suggested in \cite{Fay97} is attractive,
for among other things the bosonic part of the classical 
action Eq.~\ref{F2} coincides with the Non-Abelian Born-Infeld action
after the solution classical equation of motion 
for the $Y$-field is substituted back into Eq.~\ref{F2}.
We note that the equation of motion for the $Y$-field is
\begin{equation}
\frac{\alpha}{4}\left(
Y^{-1}[A_{\mu},A_{\nu}]^{2}Y^{-1}\right)_{ij}
+\beta \delta_{ij}=0,
\label{F4}
\end{equation} 
and its solution is
\begin{equation}
Y=\frac{1}{2}\sqrt{\frac{\alpha}{\beta}}
\sqrt{-[A_{\mu},A_{\nu}]^{2}}.
\label{F5}
\end{equation}
The matrix $Y^{ab}$ plays the role of the dynamical
variable, the elements of $Y$ can fluctuate while
it matrix size is fixed. In contrast the matrix size
in the model of \cite{Kaw97} is considered as a
dynamical variable so that the partition function
includes the summation over the matrix size.
This summation process is expected to recover
the integration over $\sqrt{g}$ [as mentioned
earlier], however a proof is not clear.

	Earlier Banks et al.~\cite{Ban97} 
proposed/conjectured the matrix model description
of M-theory\footnote{The first paper to give
the N=4 and N=16 SUSY gauge quantum mechanics
was \cite{Hal85}. The N=16 is the precursor
to the M(atrix) theory. I thank M.B.Halpern for
pointing reference \cite{Hal85} out to me.}. 
Essentially this M(atrix) theory,
as it has been dubbed, is the  large $N$ limit
of maximally supersymmetric quantum mechanics
of $U(N)$ matrices. Some of the standard wisdom
about M-theory and related topics can be found in
Ref.~\cite{Mak97}.

Our motivations in the form of comments and questions for constructing a
model based on topological/algebraic arguments that has relation
to F-theory and Matrix Model IIB and conjectures 
regarding the construction of a model
in which the notions of spacetime and noncommutative spacetime 
arise out of some underlying topological/algebraic
structure are given in \cite{Alam97}.

	This paper is organized as follows. 
The next section contains the actual construction of the 
desired Topological Matrix Model. In section three we 
comment on the relation of the Topological Matrix Model
to F-theory and Matrix Model IIB. Comments 
regarding the possible origin of the chemical
potential term in Eq.~\ref{F1} are given in section four.
Conclusions are contained in section five.

	 Recently there has been a lot of interest
in Topological Yang-Mills theory in higher dimensions
consequently overlap is expected among various works.
The work of S.~Hirano and M.~Kato \cite{Hir97} has
overlap with ours\footnote{We thank M.~Kato for
pointing out their work to us}. The detailed
work of C.~Hofman and J-S.~Park \cite{Hof97}
is also worth citing in this respect. In a
forthcoming article we would like to setup
an exact comparison between our work and that
in \cite{Hof97}.
%%%%%%%%%%%%%%%%%%%%%%%%%%%%%%%%%%%%%%%%%%%%%%%%%%%%%%%%%%%%%%% 
\section{Topological Matrix Model}
	With the motivations outlined in \cite{Alam97} 
in mind we assume for the purposes of this paper that we 
have a noncommutative spacetime \footnote{Noncommutativity of 
spacetime can be considered as a first step towards
the realization of Quantum Mach Principle [QMP] \cite{Alam84,Alam97}.
QMP could simply be stated as:~{\em If there is no field/matter
there should be no space-time}\cite{Alam84,Alam97}}. The coordinates 
of this noncommutative spacetime are taken to 
be $N \times N$ matrices $X^{\mu}$. The index $\mu$ 
takes values from 0 to
$D-1$, where $D$ is the dimension of spacetime.
The value of $D$ will be fixed in the context of
relating the topological matrix model to F-theory
and type IIB matrix model, see sections three and four. 
To be particular we take the matrices $X^{\mu}$
to represent ``instantons'' so that we are naturally
led to choose the self-dual equation \ref{F8} as
our gauge-choice. The spacetime is subject
to arbitrary deformations 
$X^{\mu}\longrightarrow X^{\mu}+\Delta X^{\mu}$.
We define the measure
\begin{equation}
|\Delta X|^2 = {\rm Tr} [ \eta_{\mu\nu}
\delta X^{\mu}\delta X^{\nu}] 
\label{F5a}
\end{equation}
$\eta_{\mu\nu}$ is the Minkowski metric with
signature $(D-1,1)$. In keeping with the
ADHM description of instantons we take the 
matrices $X^{\mu}$ to lie in the adjoint
representation of the $U(N)$ group. We note
that our fundamental degrees of freedom are
D-instantons. The instantons are fuzzy, are 
the coordinates of noncommutative spacetime and
are represented by matrices. Since $X^{\mu}$ are
subject to arbitrary deformations, see Eq.~\ref{F6d}, 
the construction of the model in the present paper 
is that of topological field theory. 
The measure in Eq.~\ref{F5a}
is by definition fuzzy since $X^{\mu}$ are
subject to arbitrary deformations, Eq.~\ref{F6d}.
The fuzziness of the measure ties in nicely
with the noncommutativity of spacetime and
the QMP \cite{Alam84,Alam97}. 

	As pointed out earlier
it is tempting to go even beyond the noncommutative
spacetime. This would imply defining the ``measure''
over some topological space and recovering the 
measure over Minkowski space in Eq.~\ref{F5a}
by some suitable reduction procedure. However
for the purposes of this paper we adhere to
the measure defined in Eq.~\ref{F5a}.

%%%%%%%%%%%%%%%%%%%%%%%%%%%%%%%%%%%%%%%%%%%%%%%% 
%\section{Topological Matrix Model}
We now turn to give the Topological Matrix Model [TMM].
As is well-known \cite{Kak91,Wit88} we can classify
topological field theories into two categories:
\begin{itemize}
\item{}Topological Models with no {\em explicit} metric
dependence. Known examples in this category include 
three-dimensional Chern-Simons theory and 2+1 gravity.
\item{}Topological models where a metric may be present
but varying the background metric does not change
the theory i.e. the theory is independent of the metric.
This class of theories is called cohomological topological
field theories [CTFT]. The metric enters CTFT through
BRST gauge fixing and thus the metric is introduced
as a gauge artifact. One of the consequences of the metric
being a gauge artifact is that the energy momentum tensor
in CTFT is BRST trivial. One can see this by noting that
the energy momentum tensor [by definition] is given by
the variation of the Lagrangian with respect to the metric
\[ T_{\mu\nu}=\frac{2}{\sqrt{g}}\frac{\delta L}
{\delta g^{\mu\nu}}\] 
and since the gauged-fixed action with its Faddeev-Popov
term can be written as $\{Q,F\}$ for some field F, the
energy momentum tensor can be written as
\[ T_{\mu\nu}=\{Q, F_{\mu\nu}\}.\]
\end{itemize}

	One procedure of constructing cohomological theories
is to postulate a gauge transformation under which the original
action is invariant. The original action is taken to be zero
or a pure topological quantity. One then gets a 
Gauge-Fixed [GF] action written as a BRST variation.

We now start with zero action in the usual manner \cite{Kak91}
\begin{equation}
L = 0 
\label{F6}
\end{equation}
and construct a cohomological model. We recall that 
when considering Topological Yang-Mills symmetry
one considers the infinitesimal transformations 
\cite{Bau88}:
\begin{eqnarray}
\delta A_{\mu} = D_{\mu}\varepsilon
+\varepsilon_{\mu}, 
\label{F6a}
\end{eqnarray}
where $A=A_{\mu}dx^{\mu}$ is the Yang-Mills field and
$D_{\mu}$ is the covariant derivative
\begin{eqnarray}
D_{\mu}\equiv \partial_{\mu}
+[A_{\mu},\;\;\;], 
\label{F6b}
\end{eqnarray}
$\varepsilon$ is the usual Yang Mills local 
parameter and $\varepsilon_{\mu}$ is a
{\em new local 1-form infinitesimal parameter}.
The action of $\delta$ on the field-strength
i.e. the two-form $F=dA+AA$ is
 \begin{eqnarray}
\delta F_{\mu\nu}= D_{[\mu}\varepsilon_{\nu]}
-[\varepsilon, F_{\mu\nu}]. 
\label{F6c}
\end{eqnarray}

	In lieu of the discussion at the beginning of
this section 
and keeping in mind Eq.~\ref{F6a} we subject the non-commutative 
coordinates to arbitrary deformations and assume
that
 \begin{eqnarray}
\delta X^{\mu}=\varepsilon^{\mu}, 
\label{F6d}
\end{eqnarray}
where $\varepsilon^{\mu}$ are $N \times N$
matrices. As pointed out earlier the subjection
of  $X^{\mu}$ to arbitrary deformations implies
that we are dealing with a topological field
theory.

%%%%%%%%%%%%%%%%%%%%%%%%%%%%%%%%
The zero action \ref{F6} is assumed to be invariant
under the gauge transformation
\begin{equation}
\delta_{1}X_{\mu}=\psi_{\mu}. 
\label{F7}
\end{equation}

Next we choose a gauge
so that $[X_{\mu},X_{\nu}]$ is self-dual 
~\cite{Cor83},
%~\cite{Cor83,Alam97}, 
\begin{eqnarray}
\lambda [X_{\mu},X_{\nu}]&=& \frac{1}{2}
S_{\mu\nu\alpha\beta}[X^{\alpha},X^{\beta}], \nonumber\\
F_{\mu\nu}&\equiv& i[X_{\mu},X_{\nu}], \nonumber\\
\lambda F_{\mu\nu}&\equiv& \frac{1}{2}
S_{\mu\nu\alpha\beta}F^{\alpha\beta}.
\label{F8}
\end{eqnarray}
in view of the motivations explained earlier.
We note that $\lambda$ is the ``eigenvalue'' in the self-dual
equation ~\cite{Cor83}. 
%~\cite{Cor83,Alam97}. 
The choice of
the gauge in Eq.~\ref{F8} is nothing but the fixing 
of the underlying topological symmetry. We recall
that underlying symmetry is topological in the
sense that $X^{\mu}$ are subject to arbitrary
deformations, see Eq.~\ref{F6d}.  
There has been a lot of interest in ``instanton''
equation, especially recently, \cite{Bau97,Ach97}. 
To this end we refer the reader to ref.~\cite{Cor83}.
%To this end we have included in the Appendix of ref.~\cite{Alam97}
%some relevant details/formulae about the higher dimensional 
%``instanton'' equation\cite{Cor83}. 

	Applying the quantization procedure to the
zero action subject to Eqs.~\ref{F6} and keeping in
mind the full BRST transformation laws including 
Eq.~\ref{F7}, viz,
\begin{eqnarray}
\delta_{1}X_{\mu}&=& \psi_{\mu}, \nonumber\\
\delta_{1}\chi_{\mu\nu}&=& i B_{\mu\nu}, \nonumber\\
\delta_{1}B_{\mu\nu} &=& 0,\nonumber\\
\delta_{1}\psi_{\mu} &=& 0,
\label{F9}
\end{eqnarray}
we may write the gauge fixed action with Faddeev-Popov [FP]
\begin{equation}
L_{_{{\rm GF+FP}}}^{1} = {\rm Tr}\left(\frac{1}{4}
B^{\mu\nu}[\lambda F_{\mu\nu}
-\frac{1}{2}S_{\mu\nu\alpha\beta}F^{\alpha\beta}]
-\chi_{\mu\nu}\left[X^{[\mu},\psi^{\nu]}\right]
+\frac{1}{8}a B^{\mu\nu}B_{\mu\nu})\right) 
\label{F10}
\end{equation}
where $\chi_{\mu\nu}$ and $\psi_{\mu}$ are the FP
ghostfields, $B_{\mu\nu}$ is a self-dual auxiliary 
field, and $a$ is a parameter which takes on in general
a different value for each component. For example
$a=a_{09}$ when $\mu=0$ and $\nu=9$ in Eq.~\ref{F10}.

We have used the subscript 1 for the BRST variation
$\delta$ to emphasize that we are carrying the quantization 
[i.e. gauge-fixing procedure] at the first stage,
in anticipation that due to hidden symmetry of the
gauge-fixed action Eq.~\ref{F10} we need to repeat
the gauge-fixing procedure.

We can write the action in Eq.~\ref{F10}
as a BRST variation using the BRST transformation 
laws given in Eq.~\ref{F9},
\begin{equation}
L_{_{{\rm GF+FP}}}^{1} =-\frac{i}{4}{\rm Tr}\left(
\delta_1(\chi^{\mu\nu}[\lambda F_{\mu\nu}
-\frac{1}{2}S_{\mu\nu\alpha\beta}F^{\alpha\beta}
+\frac{1}{2}a B_{\mu\nu}])\right). 
\label{F11}
\end{equation}

As a check we explicitly act with $\delta_{1}$ in 
Eq.~\ref{F11} and using  Eq.~\ref{F9} we arrive at
%\begin{equation}
%L_{_{{\rm GF+FP}}}^{1}= {\rm Tr}\left(
%\frac{1}{4}B^{\mu\nu}[\lambda F_{\mu\nu}
%-\frac{1}{2}S_{\mu\nu\alpha\beta}F^{\alpha\beta}]
%-\chi_{\mu\nu}\left[X^{[\mu},\psi^{\nu]}\right]
%+\frac{1}{8}a B^{\mu\nu}B_{\mu\nu})\right) 
%\label{F12}
%\end{equation}
%which is nothing but Eq.~\ref{F10}.
Eq.~\ref{F10}.

	Next we move to the second stage of gauge-fixing.
Since $\psi^{\mu}$ has a ghost-symmetry this can be 
parameterized by the ghost field $\Phi$, 
namely $\delta_{2}\psi_{\mu}=[X_{\mu},\Phi]$. 
Moreover the action given above, see Eqs.~\ref{F10}
,~\ref{F11} possesses a hidden symmetry, $\delta_{2}\psi_{\mu}=
[X_{\mu},\Phi]$, and $\delta_{2}B_{\mu\nu}= i e [\Phi,\chi_{\mu\nu}]$
where $e$ is a constant. We must thus continue the
quantization procedure by fixing this symmetry. 
To this end introduce a set of fields 
$\Phi$, $\overline{\Phi}$ and $\eta$. Keeping
these points in mind the set of BRST transformations
reads 
\begin{eqnarray}
\delta_{2}X_{\mu}&=& 0, \nonumber\\
\delta_{2}\chi_{\mu\nu}&=& 0, \nonumber\\
\delta_{2}B_{\mu\nu} &=& i e [\Phi,\chi_{\mu\nu}],\nonumber\\
\delta_{2}\psi_{\mu} &=& [X_{\mu},\Phi],\nonumber\\
\delta_{1}\Phi &=& 0 ,\nonumber\\
\delta_{2}\Phi &=& 0 ,\nonumber\\
\delta_{1}\overline{\Phi} &=& 0 ,\nonumber\\
\delta_{2}\overline{\Phi}&=& 2\eta ,\nonumber\\
\delta_{1}\eta &=& 0 ,\nonumber\\
\delta_{2}\eta &=& -\frac{1}{2}e [\Phi,\overline{\Phi}]. 
\label{F13}
\end{eqnarray}
We have used the subscript 2 for the BRST variation
$\delta$ to emphasize that we are carrying the quantization 
[i.e. gauge-fixing procedure] at the second stage.

	The gauge-fixed action subject to the BRST
rules in Eq.~\ref{F13} can be written as,
\begin{equation}
L_{_{{\rm GF+FP}}}^{2} ={\rm Tr}\left(
[\delta_1+\delta_2](-\frac{1}{2}
\overline{\Phi}[X_{\mu},\psi^{\mu}]
+\frac{1}{2}s~e\overline{\Phi}[\Phi,\eta]
+\frac{i}{4}\chi^{\mu\nu}B_{\mu\nu})\right). 
\label{F14}
\end{equation}
where $s$ is some parameter.

Carrying out explicitly the action of the BRST
variation $\delta_1+\delta_2$ in Eq.~\ref{F14}
we have,
\begin{eqnarray}
L_{_{{\rm GF+FP}}}^{2}&=&{\rm Tr}(
-\eta[X_{\mu},\psi^{\mu}]-\frac{1}{2}\overline{\Phi}
[\psi_{\mu},\psi^{\mu}]\nonumber\\
&&+\frac{1}{2}[X_{\mu},\Phi] [X^{\mu},\overline{\Phi}]
+s~e~\Phi~[\eta,\eta]\nonumber\\
&&+\frac{1}{4}s~e^2~[\Phi,\overline{\Phi}]^{2}\nonumber\\
&&-\frac{1}{4}B^{\mu\nu}B_{\mu\nu}
-\frac{1}{4}e~\Phi~[\chi^{\mu\nu},\chi_{\mu\nu}]).
 \label{F15}
\end{eqnarray}
We note that the field $\Phi$ is unaffected by
the BRST variation $\delta_1+\delta_2$. This
implies that the action Eq.~\ref{F15} is not
unique for we can add to it a BRST variation
of some fields that give a total contribution
of zero, for example we can add to the action
some arbitrary collection of the $\phi$ field
which is unaffected by the BRST variation.

The full action is the sum of the two actions
Eqs.~\ref{F10}, \ref{F15}, viz,
\begin{equation}
L_{_{{\rm GF+FP}}}=L_{_{{\rm GF+FP}}}^{1}
+L_{_{{\rm GF+FP}}}^{2}.
\label{F16}
\end{equation}

	In anticipation of comparison of TMM to other
models, we now choose the value of $D$ to be 10. This
choice is also guided by the observation that the
special properties of $\gamma$ matrices in 
eight-dimensions don't recur in higher dimensions
\cite{Cor83}. We thus choose
the self-dual equation of Eq.~\ref{F8}
  \begin{eqnarray}
\lambda [X_{\mu},X_{\nu}]&=& \frac{1}{2}
S_{\mu\nu\alpha\beta}[X^{\alpha},X^{\beta}]
\label{F8a}
\end{eqnarray}
to be valid in $D=10$ and define the totally
antisymmetric tensor $S_{\mu\nu\alpha\beta}$
\cite{Cor83},
%in analogy with Eq.~A5 in the Appendix of ref.~\cite{Alam97},
%namely\footnote{Here we use $\xi$ rather than $\eta$ of Appendix 
%of ref.~\cite{Alam97} for the constant spinor to avoid confusion
%with the $\eta$ already introduced in this note.},
%in analogy with Eq.~A5 in the Appendix of ref.~\cite{Alam97},
namely\footnote{Here we use $\xi$ rather than $\eta$
of ref.~\cite{Cor83} for the constant spinor to avoid confusion
with the $\eta$ already introduced in this note.},
\begin{equation}
S^{\mu\nu\alpha\beta}=\xi^{T}\Gamma^{\mu\nu\alpha\beta}\xi.
\label{F8b}
\end{equation}
We demand that $\xi^{T}\xi=1$ ~\cite{Cor83}.
% ~\cite{Cor83,Alam97}.
$\xi$ is a constant spinor, and $\Gamma^{\mu\nu\alpha\beta}$ 
is the totally antisymmetric product of $\Gamma$ matrices for 
$SO(9,1)$ spinor representation.
Since we want to impose 
the ``unique''  conditions \cite{Cor83}
%~\cite{Cor83,Alam97}
arising from the octonionic structure, we decompose
$\Gamma^{\mu}$ of D=10 in terms of $\gamma^{i}$ of eight-dimensional
$SO(8)$ such that 
\begin{eqnarray}
F_{0i}&=& 0, \nonumber\\
F_{9i}&=& 0, \nonumber\\
F_{09}&=& 0, i=1,2,....,8,
\label{F8c}
\end{eqnarray}
and the conditions given in Eq.~3.39 of \cite{Cor83}
hold.
%given in Eq.~A4 of the Appendix of ref.~\cite{Alam97} 
%hold \cite{Cor83}. 
Under these conditions $\lambda=1$ in Eq.~\ref{F8a}.
 The breakdown of $\Gamma^{\mu}$
in terms of $\gamma^{i}$ and the values of antisymmetric
tensor which ensures conditions given in Eq.~\ref{F8c} 
and in Eq.~3.39 of \cite{Cor83} are as follows:
\begin{eqnarray}
\Gamma^{0}&=& i\sigma_{2}\otimes 1, \nonumber\\
\Gamma^{9}&=& \sigma_{1}\otimes \gamma^{9}, \nonumber\\
\Gamma^{i}&=& \sigma_{1}\otimes \gamma^{i}, \nonumber\\
S^{0ijk}&=& 0, \nonumber\\
S^{9ijk}&=& 0, \nonumber\\
S^{09ij}&=& 0, \nonumber\\
S^{ijkl}&=& \xi^{T}\gamma^{ijkl}\xi,\;\;\;\;\;i=1,2,....,8. 
\label{F8d}
\end{eqnarray}

	Keeping in mind the information outlined
the total gauge-fixed action Eq.~\ref{F16}
can be written after integrating over the auxiliary
field $B_{ij}$ as
\begin{eqnarray}
L_{_{{\rm GF+FP}}}&=&{\rm Tr}(\frac{1}{4}F_{ij}F^{ij}
+a \frac{1}{4}F_{ij}S^{ijkl}F_{kl}\nonumber\\
&&-\frac{1}{8}F_{ij}S^{ijkl}F_{kl}
-\chi_{ij}\left[X^{[i},\psi^{j]}\right]\nonumber\\
%%%%%%%%%%%%%%%%%%%%%%%%%%
&& -\eta[X_{i},\psi^{i}]-\frac{1}{2}\overline{\Phi}
[\psi_{i},\psi^{i}]\nonumber\\
&&+\frac{1}{2}[X_{i},\Phi] [X^{i},\overline{\Phi}]
+s~e~\Phi~[\eta,\eta]\nonumber\\
&&+\frac{1}{4}s~e^2~[\Phi,\overline{\Phi}]^{2}
-\frac{1}{4}e~\Phi~[\chi^{ij},\chi_{ij}]).
 \label{F8e}
\end{eqnarray}
The set of bosonic fields in Eq.~\ref{F8e} is
$(X^{i},\Phi,\overline{\Phi})$ where the fermionic
set is $(\psi^{i},\chi^{ij},\eta)$.
The action in Eq.~\ref{F8e} is now in the form to be
compared to the supersymmetric reduced model.

	We next choose the value of $D$ to be 9,
so that we are starting with $SO(8,1)$. In order
to exploit the ``instanton'' equation in 7
Euclidean dimension \cite{Cor83}  
%\cite{Alam97} 
we consider
$SO(8,1)$ broken into $SO(1,1)\otimes SO(7)$.
Further the subgroup of $SO(7)$ which respects
the octonion structure\cite{Cor83} is $G_{2}$.
The gauge conditions in this case are obtained
by replacing 9 by 8 in Eq.~\ref{F8c}, letting $i$
run from 1 to 7, 
%\begin{eqnarray}
%F_{0i}&=& 0, \nonumber\\
%F_{8i}&=& 0, \nonumber\\
%F_{08}&=& 0, i=1,2,....,7,
%\label{F8f}
%\end{eqnarray}
and deleting the terms with subscript 8 in the
appropriate equations 
as explained in ref.~\cite{Cor83}, obtaining
the set of seven equations \cite{Cor83}.
%in Eq.~A4 of the Appendix of \cite{Alam97},
%obtaining the set of seven equations given in
%Eq.~A15 of the Appendix of \cite{Alam97}. 
We note that under these conditions $\lambda=1$ in Eq.~\ref{F8a}.

When the value of $D$ is set to 8, in a like
manner we start with  $SO(7,1)$ and consider
$SO(7,1)$ broken into $SO(1,1)\otimes SO(6)$.
The relevant subgroup in this case 
of $SO(6)$ is $SU(3)\otimes U(1)/Z_{3}$ 
\cite{Cor83}.
%\cite{Cor83,Alam97}. 
The gauge conditions in this 
case are obtained by replacing 8 by 7 \cite{Cor83},
%in Eq.~27 of \cite{Alam97}, 
letting $i$ run from 1 to 6 
%\begin{eqnarray}
%F_{0i}&=& 0, \nonumber\\
%F_{7i}&=& 0, \nonumber\\
%F_{07}&=& 0, i=1,2,....,6,
%\label{F8g}
%\end{eqnarray}
and deleting the terms with subscript 7 in the
appropriate equations 
as explained in ref.~\cite{Cor83}, obtaining
the set of seven equations \cite{Cor83}.
%in Eq.~A15 of the Appendix of \cite{Alam97},
%obtaining the set of seven equations given in
%Eq.~A17 of the Appendix of \cite{Alam97}.
Of course under these conditions $\lambda=1$ in Eq.~\ref{F8a}. 
We note that there are two subgroups of $SO(6)$
which allow an invariant construction of the
fourth rank tensor $S_{\mu\nu\alpha\beta}$
namely $SO(4)\otimes SO(2)$ and 
$SU(3)\otimes U(1)/Z_{3}$ \cite{Cor83}. The choice
$SO(4)\otimes SO(2)$ corresponds to the case
where the six dimensional manifold is a
direct product of a four dimensional and
a two dimensional manifold. The second subgroup
$SU(3)\otimes U(1)/Z_{3}$  is the holonomy
group of six dimensional Kahler manifolds.

%%%%%%%%%%%%%%%%%%%%%%%%%%%%%%%%%%%%%%%%%%%%
\section{Relationship between TMM \& other String Models}
In this section we look at the relationship
between TMM and other string theories. In particular
we want to see the relationship between TMM and 
F-Theory \cite{Vaf96}, TMM and matrix model of
M-theory and TMM and the matrix model of type 
IIB string theory\cite{Kaw97}.

	If one were to ask for a model, based on 
D-dimensional Yang-Mills type, to be written
on purely intuitive ground, the action
\begin{equation}
S=-\alpha\left(\frac{1}{4}
Tr([A_{\mu},A_{\nu}]^{2})\right),
\label{R1}
\end{equation} 
would come to mind, where $A_{\mu}$ is the Yang-Mills
field. Further insistence on incorporating supersymmetry 
would lead us to the modified action 
\begin{equation}
S_{_{SRM}}=-\alpha\left(\frac{1}{4}
Tr([A_{\mu},A_{\nu}]^{2})
+\frac{1}{2}Tr(\overline{\psi}\Gamma^{\mu}
[A_{\mu},\psi])\right),
\label{R2}
\end{equation}
where $\psi$ is a ten-dimensional Majorana-Weyl spinor
field. Eq.~\ref{R2} is nothing but the action
for supersymmetric reduced model \cite{Kaw97} and is the
same as the action in Eq.~\ref{F1} without the
$\beta N$ term. The action in Eq.~\ref{R2}
is called supersymmetric reduced model [SRM].
The ten dimensional Super Yang-Mills action
i.e. SRM of Eq.~\ref{R2}
can be rewritten in terms of octonions
of eight-dimensional space as \cite{Cor83}
%[see the Appendix of ref.~\cite{Alam97}] as
\begin{eqnarray}
S_{_{SRM}}&=&{\rm Tr}(
%%%%%%%Bosonic part
-\frac{1}{4}([A_{i},A_{j}]^{2})
+\frac{1}{2}([A_{i},A_{0}+A_{9}][A^{i},A^{0}-A^{9}])\nonumber\\
&&+\frac{1}{8}([A_{0}+A_{9},A_{0}-A_{9}]^{2})\nonumber\\
%%%%%%%%Mixed +fermi part
&&-\frac{1}{2}\lambda_{a}^{L}(2[A^{[8},\lambda^{a]}_{_R}]
+c_{abc}[A^{[b},\lambda^{c]}_{_R}])-\lambda^{8}_{_L}
[A_{i},\lambda^{i}_{_R}]\nonumber\\
&&-\frac{1}{2}(A_{0}-A_{9})[\lambda_{i}^{R},\lambda^{i}_{_R}]
-\frac{1}{2}(A_{0}+A_{9})[\lambda^{8}_{_L},\lambda^{8}_{_L}]
-\frac{1}{2}(A_{0}+A_{9})[\lambda_{a}^{L},\lambda_{a}^{L}])
\label{R3}
\end{eqnarray}
where the indices $i,j=1,.....,8$ and $a,b,c=1,...,7$ \cite{Cor83}.
%as in the Appendix of ref.~\cite{Alam97}. 
We note that the $L,R$ appearing as subscript or superscript denote 
the chirality of the $SO(8)$ chiral spinors $\lambda$. 
We have written the action of SRM in the form displayed
in Eq.~\ref{R3} to facilitate comparison with
the TMM of the previous section. To the same end
we have dropped the coupling $\alpha$. It is 
straightforward to see that the action of
Eq.~\ref{R3} is the same as the action of
TMM in $D=10$ viz Eq.~\ref{F8e}, if one lets
\begin{eqnarray}
   X^{i} &\Longleftrightarrow& A^{i},\nonumber\\
   \Phi  &\Longleftrightarrow& A_{0}+A_{9},\nonumber\\
\overline{\Phi}&\Longleftrightarrow& A_{0}-A_{9},\nonumber\\
       \psi^{i}&\Longleftrightarrow& \lambda^{i}_{_R},\nonumber\\
   2\chi^{8a}  &\Longleftrightarrow& \lambda^{a}_{_L},\nonumber\\
   \eta   &\Longleftrightarrow& \lambda^{8}_{_L},
\label{R4}
\end{eqnarray}
except for the term ${\rm Tr}(a \frac{1}{4}F_{ij}S^{ijkl}F_{kl}
-\frac{1}{8}F_{ij}S^{ijkl}F_{kl})$\footnote{The simplest gauge choice
is to take $a=0$. The trace of $F_{ij}S^{ijkl}F_{kl}$
for finite $N$ vanishes due to Jacobi-identity and the
cyclic property of trace. In large N limit this term may
survive and could play a role in the dynamics of the matrix model. 
We do not understand the implications of this term on the 
dynamics of the matrix model.}.

		In order to make connection of the TMM
with F-Theory we note that F-Theory is formulated in
12 dimensions \cite{Vaf96} and is supposed to be the
underlying theory of type IIB strings. More precisely
F-Theory is defined only through the compactifications on
elliptically fibered complex manifolds. For the purposes
of this paper we compare the TMM with F-theory in
a naive manner ignoring for the moment the task
of compactifying TMM according to complicated
compactification schemes, such as, for example 
compactification on $K_{3}$ orbifold and $T^{4}/Z_{2}$. 

	     If we look at the bosonic content of 
TMM we can interpret the emergence of $\Phi$
and $\overline{\Phi}$ as two ``extra coordinates''. 
We see that besides the 10 dimensional spacetime we have 
to contend with two ``extra dimensions'' $\Phi+\overline{\Phi}$
and $\Phi-\overline{\Phi}$. More precisely in addition
to the eight transverse coordinates $X^{i}$ 
[$i=1,2,...8.$] we have the light-cone coordinates 
$X^{0}$ and $X^{9}$ and two extra transverse 
coordinates $\Phi+\overline{\Phi}$
and $\Phi-\overline{\Phi}$. If we write the
contribution of these 12 coordinates to explicitly
show the signature we have $(1,1)$, $(8,0)$
and (1,1) coming from $X^{0}$ and $X^{9}$, $X^{i}$
and $\Phi+\overline{\Phi}$ and $\Phi-\overline{\Phi}$
respectively. Thus the TMM has 10+2 spacetime
dimensions. The lightcone TMM seems to correspond to
lightcone F-theory with 9+1=10 transverse coordinates.

	It is known \cite{Vaf96} that if we compactify
F-theory on (1,1) space we will obtain type IIB
string theory. Thus if we compactify the (1,1) space
i.e. $(\Phi+\overline{\Phi},\Phi-\overline{\Phi})$ 
we will obtain a matrix description of the
light-cone type IIB theory. Thus a matrix description
of the light-cone type IIB theory can be taken to be
the SRM on $T^{1,1}$ torus with $\Phi$ and 
$\overline{\Phi}$ directions compactified.

We now turn to the cases of D=9 and D=8.
Let us begin with the $D=9$ case, where we 
start with the 9 dimensional spacetime.
Taking into account the two extra dimensions 
[see discussion above], the TMM has 9+2 spacetime 
dimensions. If we compactify $\Phi$ and 
$\overline{\Phi}$ directions to obtain the
$T^{1,1}$ torus, we may write $R^{9,2}
\rightarrow R^{8,1}\times T^{1,1}$. Now in 
M-theory $R^{10,1}\rightarrow R^{9,1}
\times S^{1}$ \cite{Ban97}. If we compactify  $R^{9,1}$
on two torus $T^{2}$, we obtain $R^{10,1}
\rightarrow (R^{7,1} \times S^{1})\times T^{2}$.
At this point we recall that the `fundamental'
excitations of M-theory {\em ala} 
Banks et al.\cite{Ban97} are 0-branes.
In the present model the basic objects
are `instantons' [-1-branes]. We 
conjecture/expect that the TMM model 
compactified on two-torus, namely,
$R^{9,2}\rightarrow R^{8,1}\times T^{1,1}$
is equivalent to M-theory $R^{10,1}
\rightarrow (R^{7,1}\times S^{1})\times T^{2}$. 
For the case of the 8 dimensional spacetime,
when we compactify our theory on two-torus
we can write $R^{8,2}\rightarrow R^{7,1}
\times T^{1,1}$. We may obtain this case
by compactifying its higher dimensional
counterparts. In the above compactification
schemes we have used the simple conventional
logic. However as pointed out in \cite{Set97}
the conventional logic leads one to expect
that if M-theory is compactified on a two-torus,
then since 11-2=9 one should obtain a 9
dimensional theory, but one finds that if
the area of the torus is shrunk to zero
the result is 10 dimensional IIB string theory.
It would be interesting to examine if
we can manufacture an extra dimension
for the above case of $R^{8,2}\rightarrow 
R^{7,1}\times T^{1,1}$.

	Finally we comment on the question:
What can one say in the context of
TMM about the emergence of commutative
spacetime and general coordinate transformations?
We can write 
\begin{eqnarray}
X^{i} &\longrightarrow& X^{i}+\delta X^{i},\nonumber\\
\Phi  &\longrightarrow& \Phi + \delta\Phi,\nonumber\\
\overline{\Phi}  &\longrightarrow& \overline{\Phi}
+\delta\overline{\Phi}
\label{R5}
\end{eqnarray}
for the transformations of the bosonic fields.
Since $\Phi$ is unchanged under the transformations,
as mentioned earlier, see Eq.~\ref{F13}, 
we can ignore the transformation
relation of $\Phi$ in Eq.~\ref{R5}.
The machinery of recovering the commutative
spacetime of the observed world from the
noncommutative one is not built in the
present topological model. Thus for the present
we assume a background in which the matrices 
are commuting
and replace quantities in Eq.~\ref{R5}
by their commuting counterparts [i.e
$X^{i}\rightarrow x^{i}$, $\delta X^{i}
\rightarrow g^{i}(x^i,\phi,\overline{\phi})$
$\Phi \rightarrow \phi$, $\overline{\Phi} 
\rightarrow \overline{\phi}$, 
$\delta\overline{\Phi}
\rightarrow g^{\overline{\phi}}(x^i,\phi,\overline{\phi})$
where the background fields $x^{i}$, $\phi$
and $\overline{\phi}$ are all mutually
commuting] then Eq.~\ref{R5} takes the form of
the general coordinate transformations.

	The topological model may be considered as a 
preliminary first step in an
attempt to:
\begin{itemize}
\item{}Formulate a theory underlying the several
known string theories. 
\item{}Construct a theory which has background
independence built into it.
\item{}Understand the true vacuum of string
theory. 
\item{}Formulate a theory in which spacetime
is a derived concept. It is conjectured that
the primordial vacuum has a topological structure
without a metric. The metric is expected to 
arise by quantum topological fluctuation.
\item{}Attempt to understand: The details of
how the electroweak vacuum can be accounted
for in terms of the true string vacuum.   
\end{itemize}

	 It could be useful to examine the following
questions in an attempt to formulate a fundamental
unified theory of strings:
\begin{itemize}
\item{}Can we understand Matrix Models in terms
of Knots? \cite{Are97}
\item{}Is there a Generalized Uncertainty Principle 
in the context of strings? \cite{Yon97}
\item{}What formalism of strings is best suited
to identify and describe the true vacuum of
string theory?
\end{itemize}
%%%%%%%%%%%%%%%%%%%%%%%%%%%%%%%%%%%
\section{The Chemical Potential term}
We note that the action given in Eq.~\ref{R2}
does not contain the chemical potential
term. In this section we want to address the question:
How does one account for the the chemical
potential term, $\beta N$ term in Eq.~\ref{F1}? 
We now give some brief comments/conjectures
regarding the origin of the chemical potential term 
in Eq.~\ref{F1}. To this end we recall that
this term may be traced back to the 
$\sqrt{g}$ appearing in the Schild action.
\begin{itemize}
\item{}The term $\sqrt{g}d^2\sigma$ appearing
in Schild action is nothing but the area term.
The question thus arises if we can {\it consider
this term as arising from the area preserving
diffeomorphisms of F and M theories}. Indeed
it has been recently claimed by Sugawara \cite{Sug97}
that his F theory and M theory can be formulated
as gauge theories of area preserving diffeomorphisms
algebra. We note that the M-theory of Sugawara
\cite{Sug97} is 1-brane formulation rather
than the 0-brane formulation of Banks et al.
\cite{Ban97} and the F-theory of Sugawara
\cite{Sug97} is 1-brane formulation rather
than the -1-brane formulation of 
Ishibashi et al.~\cite{Kaw97}. Assuming that the 
reverse of Sugawara's suggestion is true,
we can regard the area term as arising
from those diffeomorphisms of F and M theories
which preserve the area.
\item{}In view of formulating a generalized
uncertainity principle for string theories
and keeping in mind the work of Yoneya\cite{Yon97}
we may regard the area term to be connected
with the generalized uncertainty principle. 
\item{}It is well-known from the context
of cohomological topological field theories
\cite{Kak91} that the action is not unique
in the sense that we can add to it a BRST
variation of some arbitrary collection of
fields. We may regard the $\beta N$ term
[Eq.~\ref{F1}] as representing that set of 
$[X^{\mu},X^{\nu}]$ which are proportional
to the identity. 
\item{}The chemical potential term could
arise from a term which comes from the BRST
breaking.
\end{itemize}
We expect the above approaches to the determination 
of the origin of the chemical potential
term to be interelated or equivalent.
%
%%%%%%%%%%%%%%%%%%%%%%%%%%%%%%%%%%%%%%%%%%%
\section{Conclusions}
%We have obtained a simple Topological Matrix Model.
% % statements in this section have been moved to section IV
%This model may be considered as a first step in an
%attempt to:
%\begin{itemize}
%\item{}Formulate a theory underlying the several
%known string theories. 
%\item{}Construct a theory which has background
%independence built into it.
%\item{}Understand the true vacuum of string
%theory. 
%\item{}Formulate a theory in which spacetime
%is a derived concept. It is conjectured that
%the primordial vacuum has a topological structure
%without a metric. The metric is expected to 
%arise by quantum topological fluctuation.
%\item{}Attempt to understand: The details of
%how the electroweak vacuum can be accounted
%for in terms of the true string vacuum.   
%\end{itemize}
In conclusion we have obtained a simple Topological 
Matrix Model. By construction TMM has a strong
semblance to the SRM. It can be related to the
F-theory and type IIB matrix model.
This is not surprising since it is known or
expected that some topological model
is most likely to provide an underlying theory
of strings. By construction and by direct
comparison [Eqs.~\ref{R3} \& \ref{F8e}]
TMM can be regarded to be quite 
similar to the topologically twisted form 
of the supersymmetric reduced model. 
 
%	 It would be useful to examine the following
%questions in an attempt to formulate a fundamental
%unified theory of strings:
%\begin{itemize}
%\item{}Can we understand Matrix Models in terms
%of Knots? \cite{Are97}
%\item{}Is there a Generalized Uncertainty Principle 
%in the context of strings? \cite{Yon97}
%\item{}What formalism of strings is best suited
%to identify and describe the true vacuum of
%string theory?
%\end{itemize}
%%%%%%%%%%%%%%%%%%%%%%%%%%%%%%%%%%%%%%%%%%%%%%%
\section*{Acknowledgements}
The author would like to thank H.~Kawai, K.~Hagiwara
and N.~Ishibashi for several discussions and M.~Kato 
for explaining his work. The work of the author is 
supported by JSPS fellowship of the Japanese Ministry 
of Education, Science and Culture [MONBUSHO]. 
%%%%%%%%%%%%%%%%%%%%%%%%%%%%%%%%%%%%%%%%%%%%%%%%
\\
{\bf Note Added}:

	We note that topological
theories underlying quantum mechanics on instanton
moduli spaces in matrix context are first discussed
in a recent [interesting] paper by S.~Gukov \cite{Guk97}. 
This was pointed out to me by S.~Gukov after the 
submission of this work and I thank him for this.
%%%%%%%%%%%%%%%%%%%%%%%%%%%%%%%%%%%%%%%%%%%%%%    

%%%%%%%%%%%%%%%%%%%%%%%%%%%%%%%%%%%%%%%%%%%%%%%
%\input IAsec2.tex
%\input IAsec3.tex
%\newpage
%\input IAack.tex
%%%%%%%%%%%%%%%%%%%%%%%%%References%%%%%%%%%%%%%%%%%%%%%%%%
%\input IAref1.tex
%Ins
%\begin{thebibliography}{99}

%\end{thebibliography}
%\input IAfig.tex
%\input IAapp.tex
%\newpage
%\input IAFIG1.tex
\end{document}